\documentclass[conference]{IEEEtran}
\IEEEoverridecommandlockouts

\usepackage{amsmath,amssymb,amsfonts}
\usepackage{newtxtext,newtxmath}
\usepackage{graphicx}
\usepackage[table,dvipsnames]{xcolor}
\usepackage{textcomp}
\usepackage{booktabs}
\usepackage{multirow}
\usepackage{tabularx}
\usepackage{array}
\usepackage{microtype}
\usepackage{float}
\usepackage{balance}
\usepackage[hidelinks]{hyperref}
\usepackage{tikz}
\usepackage{pgfplots}
\pgfplotsset{compat=1.18}
\usetikzlibrary{arrows.meta,positioning,shapes.geometric,fit,calc,backgrounds,
                decorations.pathreplacing,patterns}
\usepackage[most]{tcolorbox}

\definecolor{inkA}{HTML}{1F3B63}   
\definecolor{inkB}{HTML}{2E7D6B}   
\definecolor{inkC}{HTML}{B4651B}   
\definecolor{inkD}{HTML}{5E4B8B}   
\definecolor{inkE}{HTML}{A32E2E}   
\definecolor{inkG}{HTML}{5A6472}   

\hypersetup{colorlinks=true,linkcolor=inkA,citecolor=inkA,urlcolor=inkA}

\newtcolorbox{keyfinding}[1][]{%
  enhanced, breakable=false, boxrule=0.5pt, arc=2pt,
  colback=inkA!4, colframe=inkA!65,
  left=5pt, right=5pt, top=3pt, bottom=3pt,
  fonttitle=\bfseries\footnotesize\sffamily, coltitle=white,
  colbacktitle=inkA!85, attach boxed title to top left={xshift=5pt,yshift=-2pt},
  boxed title style={boxrule=0pt,arc=1pt,left=3pt,right=3pt,top=1pt,bottom=1pt},
  fontupper=\footnotesize, #1}

\newtcolorbox{contribbox}[1][]{%
  enhanced, breakable=false, boxrule=0.5pt, arc=2pt,
  colback=inkB!5, colframe=inkB!60,
  left=5pt, right=5pt, top=4pt, bottom=4pt,
  fontupper=\footnotesize, #1}

\newcommand{\rih}[1]{\par\smallskip\noindent\textbf{#1.}\enspace\ignorespaces}
\newcommand{\system}{\mbox{TSG Suggester}}

\newcommand{\treekg}{Tree\,+\,KG}

\newcommand{\aut}[4]{%
  \makebox[0.25\textwidth][c]{%
    \begin{tabular}[t]{@{}c@{}}
      {\fontsize{11}{13.2}\selectfont #1}\\[1pt]
      {\fontsize{10}{12}\selectfont \textit{#2}}\\
      {\fontsize{10}{12}\selectfont #3}\\
      {\fontsize{10}{12}\selectfont #4}
    \end{tabular}}}

\begin{document}

\title{TSG Suggester: Tree-Structured Knowledge-Graph\\
Retrieval for Troubleshooting Guide\\
Recommendation in Cloud Incident Management}

\author{%
\aut{Shawn Pan}{Microsoft}{Cambridge, USA}{shawnpan@microsoft.com}%
\aut{PavanUttej Ravva}{Microsoft}{Redmond, USA}{pravva@microsoft.com}%
\aut{Walt Williams}{Microsoft}{Sunnyvale, USA}{walwilliams@microsoft.com}%
\aut{CJ Barberan}{Microsoft}{Cambridge, USA}{cjbarberan@microsoft.com}\\[10pt]
\aut{Nutan Sahoo}{Microsoft}{Cambridge, USA}{nutansahoo@microsoft.com}%
\aut{Ziran Min}{Microsoft}{Cambridge, USA}{ziranmin@microsoft.com}%
\aut{David Gross}{Microsoft}{Redmond, USA}{davidgross@microsoft.com}%
\aut{Irene Shaffer}{Microsoft}{Cambridge, USA}{irene.shaffer@microsoft.com}%
\thanks{All authors are with the Microsoft AI Development Acceleration Program
(MAIDAP).}%
\thanks{Product, service, and internal-system identifiers are generalized
throughout this manuscript. No verbatim guide, incident, or
customer content appears anywhere in the paper, and all reported figures are
corpus-level aggregates released under the controls described in
Section~\ref{sec:privacy}.}%
}

\maketitle

\begin{abstract}
On-call engineers in large-scale cloud services work under intense time
pressure, yet locating the correct Troubleshooting Guide (TSG) for an incident
remains a largely manual, keyword-driven process---and prior empirical work
finds that guide search consumes a substantial fraction of total mitigation
time. We present \system{}, a retrieval system that recommends relevant TSGs
directly from an incident description. We evaluate five retrieval strategies
---Text-Only RAG, Image-Augmented RAG, RAPTOR, Tree-Structured Retrieval, and
our proposed \textbf{Tree + Knowledge Graph (\treekg{})}---on \textbf{314}
real-world incidents spanning \textbf{112} unique TSGs drawn from \textbf{18}
service teams on a production incident-management system. \treekg{} converts
each guide into a tree that preserves its \emph{native} section hierarchy,
attaches LLM-generated \emph{problem abstractions} to internal nodes to bridge
the solution-oriented language of guides and the problem-oriented language of
incidents, extracts a per-guide entity knowledge graph, and fuses embedding
similarity with entity-level matching at query time. \treekg{} attains
\textbf{54.78\% Top-1} and \textbf{82.48\% Top-5} accuracy, leading every
baseline at every cutoff, with a \textbf{+8.58}-point Top-1 gain over
text-only RAG. Two findings are of independent interest. First, structural
alignment dominates: methods that preserve or rebuild document structure beat
flat chunking where precise discrimination matters. Second, and contrary to our
initial hypothesis, \textbf{multimodal enrichment actively hurts}---captioning
guide screenshots and injecting the captions costs 22.64 Top-5 points relative
to the text-only baseline, because generic captions dilute embeddings rather
than sharpening them. We report error analyses for both results and give
concrete deployment guidance.
\end{abstract}


\section{Introduction}
\label{sec:introduction}

Large-scale cloud services generate incidents continuously. When one fires, an
on-call engineer (OCE) is paged and enters a race: identify what is broken,
determine why, and restore service. The institutional answer to that race is the
\emph{Troubleshooting Guide} (TSG)---a document that encodes, for a known class
of failure, the symptoms that identify it, the diagnostic queries that confirm
it, the mitigation steps that resolve it, and the escalation path when they do
not. A mature service accumulates hundreds of these. They are the difference
between a fifteen-minute mitigation and a three-hour one.

The problem is finding the right one. Empirical study of production incident
handling shows that engineers spend a substantial fraction of total mitigation
time simply \emph{locating} the applicable guide~\cite{jiang2020mitigate}, and
broader studies of cloud incident response confirm that time-to-mitigate is
dominated by diagnosis rather than by the mechanics of the fix
itself~\cite{ghosh2022fight}. Every minute spent searching is a minute of
customer impact.

\subsection{Why TSG Retrieval Is Hard}

Three properties make guide retrieval substantially harder than generic
document search.

\textbf{The problem/solution vocabulary gap.} TSGs are written by the engineers
who solved a problem, in the language of the solution: ``run the following
query against the throttling table,'' ``check the regional failover state,''
``restart the ingestion workers in the affected scale unit.'' Incidents arrive
in the language of the symptom: ``P95 latency elevated in region X,''
``customers reporting 503s on sign-in.'' These two vocabularies can share almost
no surface tokens while describing the same failure. Keyword search fails here
by construction, and dense retrieval fails more subtly---the embeddings are
genuinely far apart because the \emph{texts} are genuinely about different
things, one about a symptom and one about a procedure.

\textbf{Embedding dilution in long, multi-topic documents.} A mature TSG is
long and covers several distinct sub-problems: a root section, then branches for
each variant of the failure, each with its own diagnostics and mitigations.
Averaging that into a single document embedding produces a vector that is
close to everything and decisive about nothing. Chunking helps, but fixed-size
chunking cuts across the document's own logical boundaries, splitting a
diagnostic step from the symptom it diagnoses.

\textbf{Near-duplicate guides.} Guides are authored independently by teams that
do not read each other's documentation. Repositories therefore accumulate
guides that address overlapping or nearly identical problems, which confuses
retrieval systems for exactly the same reason it confuses engineers.

\subsection{Our Approach}

We present \system{}, which returns a ranked list of semantically relevant TSGs
for a given incident. Our central design commitment is that \emph{document
structure is signal}, and that the right structure is the one the author already
wrote rather than one imposed post hoc by clustering.

Concretely, \treekg{} parses each guide into a JSON tree that mirrors its native
section hierarchy; generates, for every internal node, an LLM \emph{problem
abstraction}---a restatement of what problem that subtree solves and under what
conditions it triggers---which converts solution-oriented text into
problem-oriented text that lands in the same embedding neighborhood as incident
descriptions; extracts a per-guide knowledge graph of entities (services, error
codes, tools) and typed relations; and at query time fuses tree-node embedding
similarity with entity-level graph matching into a single ranking score.

We situate this against four baselines spanning the design space: flat text-only
RAG, image-augmented RAG using vision-language captioning, RAPTOR's recursive
abstractive tree~\cite{sarthi2024raptor}, and tree-structured retrieval without
the knowledge graph. We organize the study around three research questions:

\begin{contribbox}
\textbf{RQ1.} Does preserving a guide's \emph{native} hierarchy outperform flat
chunking and outperform an LLM-\emph{constructed} hierarchy such as
RAPTOR's?\\[2pt]
\textbf{RQ2.} Does adding per-guide entity-level knowledge-graph matching on
top of structure-aware embeddings improve discrimination, and where in the
ranking does the gain appear?\\[2pt]
\textbf{RQ3.} Do TSG screenshots, diagrams, and query-output images carry
retrievable signal that vision-language captioning can unlock?
\end{contribbox}

\subsection{Contributions}

\begin{itemize}
  \item A \textbf{comprehensive experimental comparison} of five retrieval
        strategies---text-only, multimodal, recursive-abstractive, native-%
        hierarchical, and knowledge-graph-augmented---on 314 real-world
        incidents and 112 guides from 18 service teams.
  \item The \textbf{\treekg{} method}, which pairs native-hierarchy trees
        carrying LLM problem abstractions with per-guide entity knowledge
        graphs, and fuses embedding similarity with entity matching at query
        time. \treekg{} leads at every cutoff, with the largest margin at
        Top-1, where discrimination is hardest.
  \item A \textbf{systematic negative result on visual understanding}. Despite
        the prevalence of screenshots and diagrams in guides, image-augmented
        retrieval falls 22.64 Top-5 points below the text-only baseline. We
        provide an error analysis identifying three distinct causes and argue
        this is a property of the retrieval stage specifically, not of
        multimodal understanding in general.
  \item \textbf{Deployment guidance} grounded in the results: return five
        candidates rather than one, invest the one-time offline KG construction
        cost, and use the system's own similarity structure to find and
        consolidate near-duplicate guides.
\end{itemize}

\section{Background and Problem Setting}
\label{sec:background}

\subsection{Anatomy of a Troubleshooting Guide}

A TSG in our corpus is a semi-structured document, authored in Markdown or HTML
by the owning service team, with a recognizable internal shape: a problem
statement, a symptom list, a set of diagnostic steps (typically telemetry
queries with expected outputs), mitigation procedures, and an escalation
section. That shape is expressed as headings and nesting, and it is the closest
thing the corpus has to a schema.

Guides also contain a significant amount of non-textual content: screenshots of
dashboards in a failing state, architecture and call-flow diagrams, and pasted
visualizations of query output. Roughly speaking, these images exist because the
author found them faster to paste than to describe, which is precisely why they
often carry information the surrounding prose does not. This is what motivated
our image-augmented branch, and Section~\ref{sec:res_visual} reports what
happened when we tested it.

\subsection{The Retrieval Setting}

At incident time, the OCE has an incident title, a severity, an
automatically generated summary, and a growing discussion log. They need a
short ranked list of candidate guides. Two properties of the setting shape the
system design. First, \emph{the user is under time pressure and will scan a
handful of candidates but not fifty}, which makes Top-1 through Top-5 the
metrics that matter and makes deep-ranking metrics largely irrelevant. Second,
\emph{a wrong-but-plausible suggestion is expensive}, because following an
inapplicable guide costs minutes of mitigation time; this argues for returning a
small set with honest ranking rather than a single confident answer.

\section{Related Work}
\label{sec:related}

\subsection{Incident Management and AIOps}

Automated incident management has grown into an active area. Empirical studies
characterize how production incidents are actually handled, where mitigation
time is spent, and what fraction of incidents are recurrences of known
failures~\cite{ghosh2022fight,chen2020linked}, and broader surveys map the AIOps
failure-management landscape~\cite{notaro2021aiops}. A large recent thread
applies LLMs to root-cause analysis and mitigation recommendation: Ahmed et
al.~\cite{ahmed2023recommending} generate root causes and mitigation steps from
incident metadata; Chen et al.~\cite{chen2024rcacopilot} build a retrieval-based
root-cause copilot that reasons over historically similar incidents; Zhang et
al.~\cite{zhang2024incontext} show that in-context learning with a frontier
model can match fine-tuned pipelines for automated root causing; and Jin et
al.~\cite{jin2023assess} use LLMs to assess and summarize outages.

These systems answer \emph{what went wrong}. \system{} answers the
complementary and logically prior question of \emph{which existing procedure
applies}, which matters because a large share of incidents are recurrences for
which a correct procedure already exists and merely needs to be found.

\subsection{Troubleshooting Guide Recommendation}

The most directly related prior work is Jiang et
al.~\cite{jiang2020mitigate}, who establish empirically that guide search is a
major component of mitigation time and propose a deep learning recommender that
matches incidents to guides. More recently, An et al.~\cite{an2024nissist} build
a multi-agent copilot that extracts actionable steps from unstructured guides
and proactively proposes mitigation plans during live incidents.

\system{} differs in where it places the intelligence. Prior recommenders treat
each guide as an atomic unit to be matched, and prior copilots treat the guide
as a source of executable steps once it has been selected. We instead argue that
the guide's \emph{internal structure} is the retrieval signal: decomposing a
guide into its native hierarchy and rewriting each subtree into problem-oriented
language changes what is being matched, and doing so is what closes the
vocabulary gap that makes atomic matching hard. The knowledge graph then
supplies a complementary discrete signal for cases where the continuous one is
ambiguous.

\subsection{Dense Retrieval and Retrieval-Augmented Generation}

Modern retrieval rests on dense bi-encoders trained with contrastive
objectives---DPR~\cite{karpukhin2020dpr}, Contriever~\cite{izacard2022contriever},
and general-purpose embedding families such as E5~\cite{wang2022e5} and
BGE~\cite{xiao2023cpack}---together with late-interaction models that retain
token-level granularity~\cite{khattab2020colbert,santhanam2022colbertv2}, and
the sparse BM25 baseline that remains stubbornly
competitive~\cite{robertson2009bm25}. Sentence-BERT~\cite{reimers2019sbert}
established the practical recipe for producing semantically meaningful
fixed-size sentence vectors and is the embedding model in our pipeline.
Retrieval-augmented generation~\cite{lewis2020rag} popularized grounding
generation in retrieved evidence, and the design space has since been
systematized in surveys~\cite{gao2023ragsurvey}.

Our contribution is orthogonal to advances in the encoder. We hold the embedding
model fixed across all five methods and vary only \emph{what text gets embedded
and how it is organized}, which isolates the effect of representation structure
from the effect of encoder quality.

\subsection{Hierarchical and Structure-Aware Retrieval}

The observation that flat chunking discards document organization has produced
several responses. RAPTOR~\cite{sarthi2024raptor} recursively clusters and
summarizes chunks bottom-up into a multi-level tree, so retrieval can match at
several levels of abstraction; this is a \emph{constructed} hierarchy, imposed
by clustering rather than read from the document. Late
chunking~\cite{gunther2024latechunking} takes a different route, embedding the
full document with a long-context model before pooling into chunks so that each
chunk embedding retains global context.

Most relevant to us is BookRAG~\cite{wang2025bookrag}, concurrent work that
targets documents with a native hierarchical structure such as books,
handbooks, and technical manuals. BookRAG builds a compound index---\emph{
BookIndex}---that pairs a hierarchical tree extracted from the document (playing
the role of its table of contents) with an entity graph capturing relations
among entities mentioned within that document, plus a mapping from entities to
tree nodes. At query time an agent-based router inspired by Information Foraging
Theory classifies each query and dispatches it to a tailored retrieval workflow,
either local tree traversal or global multi-hop graph reasoning. BookRAG reports
state-of-the-art retrieval recall and QA accuracy on three benchmarks relative
to flat-chunk baselines.

The convergence is striking and, we think, informative: BookRAG and \treekg{}
independently arrive at the same three-part recipe---native hierarchy, per-%
document entity graph, and an explicit link between the two---from different
starting points and for different document classes. The differences are
instructive as well. BookRAG targets general question answering over books and
manuals and routes queries adaptively at the agent level; \system{} targets a
single retrieval task in a domain with a specific and severe vocabulary
mismatch, and it addresses that mismatch with a component BookRAG does not
require: LLM-generated \emph{problem abstractions} attached to internal nodes,
which rewrite solution-oriented content into the problem-oriented register of
the query. We regard the independent convergence as evidence that
structure-plus-entity indexing is the right general shape for retrieval over
authored, hierarchical technical documents, and our negative result on
multimodal enrichment (Section~\ref{sec:res_visual}) as a domain-specific
caution that structure should be reinforced rather than diluted.

\subsection{Knowledge-Graph-Augmented Retrieval}

Graph-based retrieval augments continuous similarity with discrete relational
structure. GraphRAG~\cite{edge2024graphrag} extracts an entity graph over a
corpus and uses community detection to build hierarchical summaries that support
global, thematic queries flat RAG cannot serve.
LightRAG~\cite{guo2024lightrag} indexes a corpus as an entity-relation graph and
retrieves at two levels of specificity with incremental index updates.
HippoRAG~\cite{gutierrez2024hipporag} builds a schemaless entity graph and uses
Personalized PageRank to achieve multi-hop association in a single retrieval
step. G-Retriever~\cite{he2024gretriever} extracts a query-relevant subgraph via
a Prize-Collecting Steiner Tree formulation before handing it to an LLM, and
Think-on-Graph~\cite{sun2024tog} has the LLM itself beam-search over graph hops.
The area is surveyed in~\cite{peng2024graphragsurvey}.

Nearly all of this work builds \emph{one graph over a corpus} to enable
cross-document, multi-hop reasoning. We do the opposite: we build one small,
independent graph \emph{per guide} and never merge them. This is a deliberate
fit to the task. Our retrieval unit is the guide, so cross-guide edges would
blur exactly the boundary we need to keep sharp; per-guide graphs also keep
construction embarrassingly parallel and make incremental re-indexing of a
single edited guide trivial. The cost is that we cannot answer questions
requiring reasoning across guides, which is not the task. We return to merging
in Section~\ref{sec:future}.

\subsection{Multimodal Document Understanding and Retrieval}

Document AI has produced strong models for understanding page images:
layout-aware pre-trained encoders~\cite{huang2022layoutlmv3}, OCR-free
end-to-end architectures~\cite{kim2022donut}, and benchmarks establishing the
difficulty of visual document question answering~\cite{mathew2021docvqa}. For
retrieval specifically, ColPali~\cite{faysse2024colpali} embeds page images
directly with a vision-language model and matches with late interaction,
bypassing text extraction entirely and outperforming OCR pipelines on visually
complex documents; M3DocRAG~\cite{cho2024m3docrag} combines visual page
retrieval with a VLM reader for multi-page, multi-document question answering.

Our image-augmented baseline follows the pragmatic, widely used alternative:
caption images with a VLM and inject the captions as text. This is the approach
most teams can deploy without replacing their retrieval stack, and it is
therefore the one worth testing honestly. Our result---that it \emph{hurts}
substantially---is not a refutation of ColPali-style native visual retrieval,
which never reduces the image to a caption. It is a warning about the
caption-and-inject pattern specifically, and Section~\ref{sec:analysis_visual}
explains the mechanism.

\subsection{Evaluation}

We report Top-$k$ accuracy against ground-truth guide labels, which is the
metric the deployment setting dictates. We note for completeness that the
broader literature has moved toward judge-based and reference-free evaluation of
retrieval-augmented systems~\cite{zheng2023judging,es2024ragas,saadfalcon2024ares},
which becomes necessary when there is no single correct answer. Section
\ref{sec:future} discusses why that is the right direction for this task too,
given that our ground-truth labels record the guide an engineer \emph{used},
not necessarily the best guide available.

\section{Problem Formulation}
\label{sec:problem}

Let an incident $I$ be described by its title, severity, automatically generated
summary, and discussion log, and let $\mathcal{T} = \{T_1, T_2, \ldots, T_n\}$
be a corpus of troubleshooting guides. The retrieval task is to produce a ranked
list $\hat{\mathcal{T}}_k \subset \mathcal{T}$, $|\hat{\mathcal{T}}_k| = k$, of
the guides most likely to apply to $I$.

We evaluate with \textbf{Top-$k$ accuracy}: the fraction of incidents for which
the ground-truth guide $T^{\star}(I)$ appears in $\hat{\mathcal{T}}_k$, for
$k \in \{1,3,5\}$:
\[
  \mathrm{Acc}@k \;=\; \frac{1}{|\mathcal{I}|}
  \sum_{I \in \mathcal{I}} \mathbb{1}\!\left[\,T^{\star}(I) \in
  \hat{\mathcal{T}}_k(I)\,\right].
\]
Ground truth is the guide recorded against the incident in the
incident-management system by the engineer who resolved it.

\section{Methodology}
\label{sec:methodology}

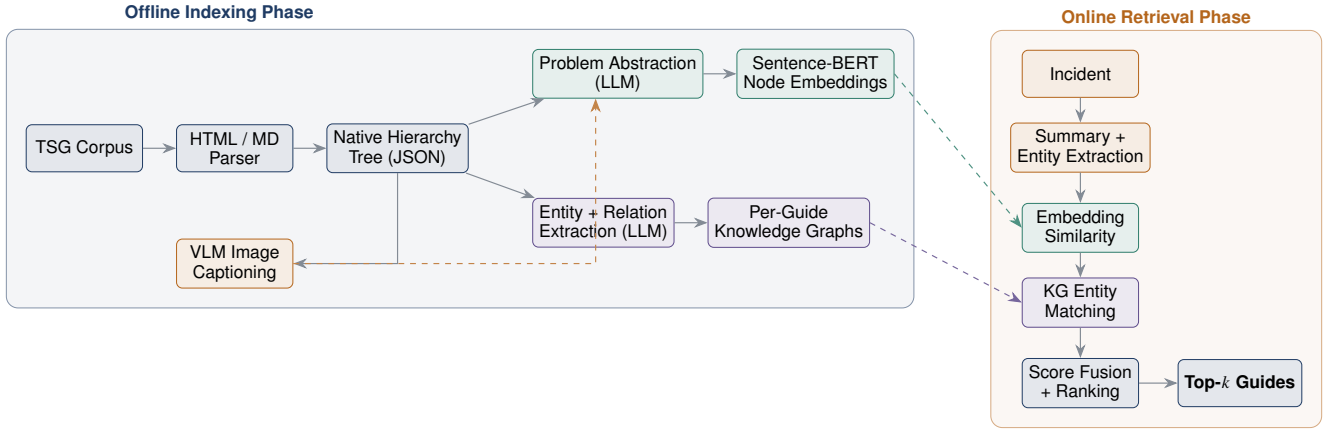
\begin{figure*}[t]
\centering
\resizebox{0.96\textwidth}{!}{%
\begin{tikzpicture}[
    node distance=0.5cm and 0.6cm,
    >={Stealth[length=2.5mm]},
    box/.style={rectangle, rounded corners=3pt, draw=#1, fill=#1!12,
                minimum height=0.85cm, minimum width=2.1cm,
                align=center, font=\small\sffamily},
    phase/.style={rectangle, rounded corners=6pt, draw=#1!55, fill=#1!5,
                  inner sep=10pt},
    lbl/.style={font=\small\sffamily\bfseries, color=#1},
    arr/.style={->, semithick, color=inkG!75},
]

\node[box=inkA] (tsg) {TSG Corpus};
\node[box=inkA, right=of tsg] (parse) {HTML / MD\\Parser};
\node[box=inkA, right=of parse] (tree) {Native Hierarchy\\Tree (JSON)};

\node[box=inkB, above right=0.45cm and 1.15cm of tree] (abstract)
  {Problem Abstraction\\(LLM)};
\node[box=inkB, right=of abstract] (embed) {Sentence-BERT\\Node Embeddings};

\node[box=inkD, below right=0.45cm and 1.15cm of tree] (kg)
  {Entity + Relation\\Extraction (LLM)};
\node[box=inkD, right=of kg] (kgstore) {Per-Guide\\Knowledge Graphs};

\node[box=inkC, below=1.2cm of parse] (img) {VLM Image\\Captioning};

\begin{scope}[on background layer]
  \node[phase=inkA, fit=(tsg)(parse)(tree)(abstract)(embed)(kg)(kgstore)(img),
        label={[lbl=inkA]above left:Offline Indexing Phase}] (offbox) {};
\end{scope}

\draw[arr] (tsg) -- (parse);
\draw[arr] (parse) -- (tree);
\draw[arr] (tree) -- (abstract);
\draw[arr] (tree) -- (kg);
\draw[arr] (abstract) -- (embed);
\draw[arr] (kg) -- (kgstore);
\draw[arr] (tree) |- (img);
\draw[arr, dashed, inkC!80] (img.east) -| ([xshift=-4mm]abstract.south);

\node[box=inkC, right=2.3cm of embed] (incident) {Incident};
\node[box=inkC, below=0.45cm of incident] (aisumm)
  {Summary +\\Entity Extraction};
\node[box=inkB, below=0.55cm of aisumm] (embsim) {Embedding\\Similarity};
\node[box=inkD, below=0.45cm of embsim] (kgmatch) {KG Entity\\Matching};
\node[box=inkA, below=0.55cm of kgmatch] (fuse) {Score Fusion\\+ Ranking};
\node[box=inkA, right=0.7cm of fuse] (topk) {\textbf{Top-$k$ Guides}};

\begin{scope}[on background layer]
  \node[phase=inkC, fit=(incident)(aisumm)(embsim)(kgmatch)(fuse)(topk),
        label={[lbl=inkC]above:Online Retrieval Phase}] {};
\end{scope}

\draw[arr] (incident) -- (aisumm);
\draw[arr] (aisumm) -- (embsim);
\draw[arr] (embsim) -- (kgmatch);
\draw[arr] (kgmatch) -- (fuse);
\draw[arr] (fuse) -- (topk);

\draw[arr, dashed, inkB!85] (embed.east) -- (embsim.west);
\draw[arr, dashed, inkD!85] (kgstore.east) -- (kgmatch.west);

\end{tikzpicture}%
}
\caption{End-to-end \system{} architecture. The \emph{offline indexing phase}
(left) parses each guide into a tree mirroring its native section hierarchy,
generates a problem abstraction and a knowledge graph for every subtree, and
embeds the abstractions. The \emph{online retrieval phase} (right) summarizes
the incoming incident, extracts its entities, and ranks guides by fusing
tree-node embedding similarity with per-guide knowledge-graph entity matching.
The image-captioning branch (amber) is evaluated as a baseline and is
\emph{not} part of the final \treekg{} configuration---see
Section~\ref{sec:res_visual}.}
\label{fig:overview}
\end{figure*}

Figure~\ref{fig:overview} presents the end-to-end architecture. The system has
an \emph{offline indexing phase} that converts each guide into tree-structured
embeddings plus a per-guide knowledge graph, and an \emph{online retrieval
phase} that scores incoming incidents against the index using a hybrid of
embedding similarity and entity matching.

\subsection{Dataset}
\label{sec:dataset}

We constructed a real-world evaluation dataset from a production
incident-management system. A telemetry query retrieved over 30,000 incidents
from production services; after filtering to incidents with an associated
ground-truth guide, the final dataset comprises \textbf{112 unique TSGs} and
\textbf{314 sampled incidents} drawn from \textbf{18 service teams}
(Table~\ref{tab:dataset}). Each incident carries its title, severity,
automatically generated summary, and chronological discussion log. Each guide
contains structured sections---problem description, symptoms, diagnostic steps,
mitigation procedures---and may embed images such as dashboard screenshots,
call-flow diagrams, and query-output visualizations.

\begin{table}[H]
\centering
\caption{Evaluation corpus statistics. Service and team identities are
generalized; the incident fields and guide content used are described in the
text above.}
\label{tab:dataset}
\footnotesize
\setlength{\tabcolsep}{5pt}
\begin{tabular}{@{}lr@{}}
\toprule
\textbf{Corpus property} & \textbf{Value} \\
\midrule
Incidents retrieved (pre-filter)        & $>$30{,}000 \\
Incidents evaluated (with ground truth) & 314 \\
Unique troubleshooting guides           & 112 \\
Service teams represented               & 18 \\
\midrule
Incidents per guide (mean)              & 2.8 \\
\bottomrule
\end{tabular}
\end{table}

\subsection{Baseline Methods}
\label{sec:baselines}

\rih{Text-Only RAG}
The flat baseline chunks each guide into fixed-size text segments, embeds them
with Sentence-BERT~\cite{reimers2019sbert}, and indexes them in a vector store.
At query time the incident description is embedded and guides are ranked by
maximum chunk cosine similarity.

\rih{Image-Augmented RAG}
Many guides contain screenshots, error dialogs, and query-output images that are
invisible to text-only embeddings. This variant uses a vision-language model to
convert each image into a textual description, injects those descriptions into
the corresponding guide sections, and then applies the identical chunking and
embedding pipeline as the text-only baseline. It is therefore a controlled test
of \emph{added visual signal}, holding everything else constant.

\rih{RAPTOR}
Following RAPTOR~\cite{sarthi2024raptor}, we recursively cluster guide chunks
and generate LLM summaries at each level, producing a tree of increasingly
abstract representations, and retrieve across all levels. The contrast with our
method is precise and deliberate: RAPTOR builds a hierarchy by clustering
content, whereas we read the hierarchy the author wrote.

\subsection{Tree-Structured Retrieval}
\label{sec:tree}

Tree-Structured Retrieval preserves the \emph{native} document hierarchy. Each
guide is parsed into a JSON tree in which sections map to nodes and subsections
to children. Images are replaced with LLM-generated textual descriptions so that
all content is unified into a single modality.

The key component is \emph{problem abstraction}. For each internal node, the
system aggregates all descendant content and prompts an LLM to emit a concise
problem description together with a set of triggering conditions. Because guides
are written in solution-oriented language (``run this query against the
throttling table'') while incidents are phrased as problems (``latency spike on
service X''), these generated abstractions bridge the vocabulary gap directly:
they produce node embeddings that live in the same region of embedding space as
incident descriptions, rather than in the region occupied by procedures.

This is the single most important design decision in the system, and it is what
distinguishes structure-aware retrieval that works from structure-aware
retrieval that merely reorganizes the same unhelpful text.

\subsection{Tree + Knowledge Graph}
\label{sec:treekg}

\begin{figure}[t]
\centering
\resizebox{\columnwidth}{!}{%
\begin{tikzpicture}[
    >={Stealth[length=2mm]},
    stage/.style={rectangle, rounded corners=3pt, draw=inkA!70, fill=inkA!10,
                  minimum height=1.65cm, minimum width=2.15cm,
                  align=center, font=\scriptsize\sffamily},
    arr/.style={->, semithick, color=inkA!70},
    node distance=0.42cm,
    lbl/.style={font=\tiny\sffamily, color=inkG, align=center},
]
\node[stage] (s1) {\textbf{Stage 1}\\Tree\\Construction};
\node[stage, right=of s1] (s2) {\textbf{Stage 2}\\Problem\\Abstraction};
\node[stage, right=of s2] (s3) {\textbf{Stage 3}\\KG\\Construction};
\node[stage, right=of s3, draw=inkC!75, fill=inkC!12] (s4)
  {\textbf{Stage 4}\\Hybrid\\Retrieval};

\draw[arr] (s1) -- (s2);
\draw[arr] (s2) -- (s3);
\draw[arr, color=inkC!70] (s3) -- (s4);

\node[lbl, below=0.15cm of s1] {guide $\to$ JSON tree\\images $\to$ text};
\node[lbl, below=0.15cm of s2] {LLM emits problem\\+ trigger conditions};
\node[lbl, below=0.15cm of s3] {LLM extracts\\entities + relations};
\node[lbl, below=0.15cm of s4] {embed.\ sim.\ $\oplus$\\entity overlap $\to$ rank};

\draw[decorate, decoration={brace, amplitude=4pt, mirror}, inkG!55]
  ([yshift=-0.95cm]s1.south west) -- ([yshift=-0.95cm]s3.south east)
  node[midway, below=4pt, font=\tiny\sffamily, color=inkG]
  {Offline, once per guide};
\draw[decorate, decoration={brace, amplitude=4pt, mirror}, inkG!55]
  ([yshift=-0.95cm]s4.south west) -- ([yshift=-0.95cm]s4.south east)
  node[midway, below=4pt, font=\tiny\sffamily, color=inkG] {Online};
\end{tikzpicture}%
}
\caption{The four-stage \treekg{} pipeline. Stages~1--3 run once per guide
offline; only Stage~4 is on the incident-response critical path.}
\label{fig:pipeline}
\end{figure}
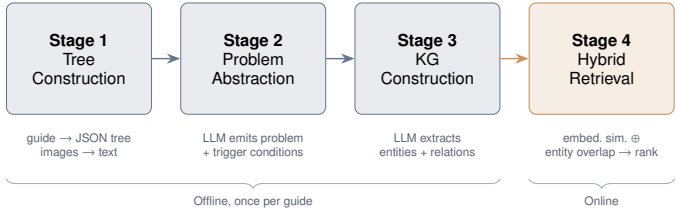

Our proposed method extends tree-structured retrieval with per-guide knowledge
graphs through a four-stage pipeline (Figure~\ref{fig:pipeline}).

\rih{Stage 1 --- Tree construction}
Each guide is parsed into a hierarchical JSON tree mirroring its native section
structure, with embedded images replaced by VLM-generated text descriptions.

\rih{Stage 2 --- Problem abstraction}
Descendant content under each internal node is aggregated and an LLM generates a
problem description and triggering conditions, rewriting solution-oriented text
into the problem-oriented register of incident descriptions
(Section~\ref{sec:tree}).

\rih{Stage 3 --- Knowledge graph construction}
The same tree is processed by an LLM to extract entities---service names, error
codes, diagnostic tools---and typed relations such as \textsf{causes},
\textsf{mitigates}, and \textsf{diagnosed\_by}, yielding one independent
knowledge graph per guide, stored as JSON. Graphs are never merged across
guides.

\rih{Stage 4 --- Hybrid retrieval}
At query time the system generates a summary of the incident, extracts its
entity set $E_I$, and scores each guide $T$ by fusing the two signals. Writing
$\mathrm{nodes}(T)$ for the abstraction-bearing nodes of $T$'s tree, $e(\cdot)$
for the embedding function, and $E_T$ for $T$'s knowledge-graph entity set:
\[
  \mathrm{score}(I,T)
  \;=\; \alpha \cdot \!\!\max_{n \in \mathrm{nodes}(T)}\!\!
        \cos\!\bigl(e(I), e(n)\bigr)
  \;+\; (1-\alpha)\cdot m\bigl(E_I, E_T\bigr),
\]
where $m(\cdot,\cdot)$ is a normalized entity-overlap score and
$\alpha \in [0,1]$ balances continuous against discrete evidence. The
$\max$ over nodes is what defeats embedding dilution: a guide is scored by its
\emph{best-matching sub-problem}, not by an average over all the sub-problems it
happens to also cover.

The two signals fail in different ways, which is why fusing them helps.
Embedding similarity is robust to paraphrase but degrades when two guides
describe similar-sounding problems in similar language. Entity matching is
brittle to paraphrase but decisive when an incident names a specific error code
or service that appears in exactly one guide. Section~\ref{sec:res_topk} shows
the gain concentrates at Top-1, which is precisely where the embedding margin
between competing candidates is smallest.

\subsection{Visual Understanding Pipeline}
\label{sec:visual}

Guides contain screenshots, flow diagrams, and query-output images that plausibly
carry diagnostic information absent from surrounding prose. We built a
multimodal pipeline that extracts images, captions them with a VLM, and injects
the captions into the appropriate guide section before embedding. Early
experiments on synthetic data were encouraging. Section~\ref{sec:res_visual}
reports what happened on real data.

\section{Data Handling and Disclosure}
\label{sec:privacy}

Troubleshooting guides and incident records are commercially sensitive and
contain personal data: guides embed internal hostnames, service topology, and
authored-by attribution; incident discussion logs contain identified employee
speech and customer or tenant identifiers. We therefore state how this study
handled that data, both because it constrains what we can report and because it
is a precondition for reproducing this work in another organization.

\rih{Scope-limited processing under existing access control}
Indexing and evaluation ran inside the environment that already holds the data,
under existing role-based access control. No guide or incident content was
exported, and no content was shared with any third-party system outside that
boundary.

\rih{Inference-only model use}
All LLM and VLM components are frozen, instruction-tuned models accessed through
inference-only endpoints with no training retention. Guide and incident content
is never used to update model weights. This matters because language models
demonstrably memorize and can regurgitate training
data~\cite{carlini2021extracting,carlini2023quantifying,huang2022leaking}.

\rih{Aggregate-only reporting}
Every number in this paper is a corpus-level aggregate over 314 incidents and
112 guides. We report no per-incident results, no per-team breakdowns, and no
per-guide accuracies. Readers should treat differences smaller than the
reported precision as not meaningful.

\rih{Manuscript-level generalization}
Product, service, team, and internal-system names are replaced
throughout with functional descriptors. No verbatim guide text, incident text,
telemetry query, or image appears anywhere in this paper, and all illustrative
phrasings are synthetic constructions matching the \emph{form} of real content
without reproducing any instance of it.

\section{Experimental Setup}
\label{sec:setup}

All five methods were evaluated on the same 314-incident dataset using Top-$k$
accuracy for $k \in \{1,3,5\}$. For each incident the system returns the top-$k$
ranked guides, and we check whether the ground-truth guide recorded in the
incident-management system appears in the returned set.

We hold the retrieval substrate constant across methods to isolate the effect of
representation. All text embeddings use Sentence-BERT~\cite{reimers2019sbert}.
A single frontier commercial LLM (GPT-4o class) performs problem abstraction,
entity and relation extraction, image captioning, and incident summarization.
Knowledge graphs are stored as per-guide JSON. Beyond aggregate accuracy we
analyzed the rank position of the correct guide in the full ranked list,
confusion patterns between semantically similar guides, and performance
stratified by incident severity and guide length.

\section{Results}
\label{sec:results}

\begin{figure}[t]
\centering
\resizebox{\columnwidth}{!}{%
\begin{tikzpicture}
\begin{axis}[
    ybar, width=8.6cm, height=5.6cm,
    bar width=7.5pt,
    ymin=0, ymax=95,
    ylabel={\footnotesize Accuracy (\%)},
    ylabel style={font=\footnotesize},
    ytick={0,20,40,60,80},
    yticklabel style={font=\scriptsize},
    symbolic x coords={treekg,tree,text,raptor,image},
    xtick=data,
    xticklabels={{\treekg{}\\(ours)},{Tree\\Structure},{Text-Only\\RAG},{RAPTOR},{Image-Aug.\\RAG}},
    xticklabel style={font=\scriptsize, align=center},
    enlarge x limits=0.13,
    axis x line*=bottom, axis y line*=left,
    ymajorgrids, grid style={inkG!18},
    nodes near coords, nodes near coords style={font=\tiny, /pgf/number format/fixed,
        /pgf/number format/precision=1},
    legend style={at={(0.5,1.03)}, anchor=south, legend columns=3,
                  font=\scriptsize, draw=none, fill=none, column sep=8pt},
]
\addplot[draw=inkA!85, fill=inkA!70] coordinates
  {(treekg,54.78) (tree,48.35) (text,46.20) (raptor,37.66) (image,27.63)};
\addlegendentry{Top-1}
\addplot[draw=inkA!70, fill=inkA!40] coordinates
  {(treekg,77.07) (tree,70.57) (text,70.40) (raptor,61.69) (image,50.45)};
\addlegendentry{Top-3}
\addplot[draw=inkA!55, fill=inkA!18] coordinates
  {(treekg,82.48) (tree,75.68) (text,81.20) (raptor,72.08) (image,58.56)};
\addlegendentry{Top-5}
\end{axis}
\end{tikzpicture}%
}
\caption{Top-$k$ retrieval accuracy across all five methods on 314 real-world
incidents. \treekg{} leads at every cutoff; the margin is largest at Top-1,
where discrimination between competing candidates is hardest.}
\label{fig:topk}
\end{figure}
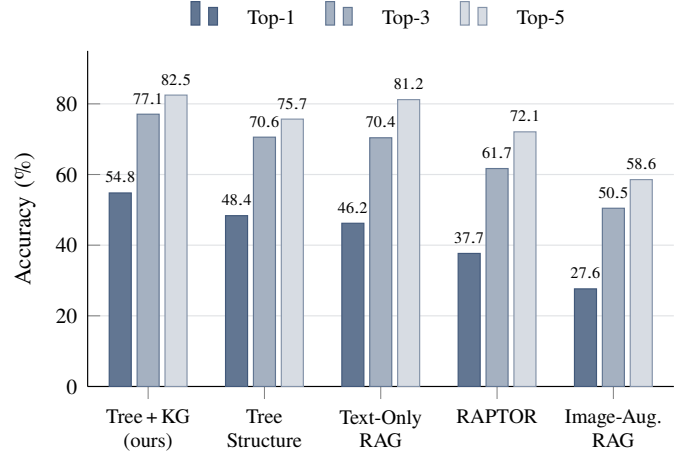

Table~\ref{tab:results} and Figure~\ref{fig:topk} present retrieval accuracy for
all five methods.

\begin{table}[t]
\centering
\caption{Top-$k$ TSG retrieval accuracy (\%) on real-world incident data
(314 incidents, 112 guides, 18 service teams). Best result per column in bold.}
\label{tab:results}
\footnotesize
\setlength{\tabcolsep}{6pt}
\begin{tabular}{@{}l ccc@{}}
\toprule
\textbf{Method} & \textbf{Top-1} & \textbf{Top-3} & \textbf{Top-5} \\
\midrule
\rowcolor{inkA!10}
\textbf{\treekg{} (ours)} & \textbf{54.78} & \textbf{77.07} & \textbf{82.48} \\
Tree Structure            & 48.35 & 70.57 & 75.68 \\
Text-Only RAG             & 46.20 & 70.40 & 81.20 \\
RAPTOR                    & 37.66 & 61.69 & 72.08 \\
Image-Augmented RAG       & 27.63 & 50.45 & 58.56 \\
\bottomrule
\end{tabular}
\end{table}

\subsection{\treekg{} Leads at Every Cutoff (RQ1, RQ2)}
\label{sec:res_topk}

\begin{figure}[t]
\centering
\resizebox{0.92\columnwidth}{!}{%
\begin{tikzpicture}
\begin{axis}[
    xbar, width=7.4cm, height=4.3cm,
    xmin=-23, xmax=11,
    xlabel={\footnotesize Top-1 accuracy $\Delta$ vs.\ Text-Only RAG (points)},
    xlabel style={font=\footnotesize},
    xtick={-20,-15,-10,-5,0,5,10},
    xticklabel style={font=\scriptsize},
    symbolic y coords={image,raptor,tree,treekg},
    ytick={image,raptor,tree,treekg},
    yticklabels={{Image-Aug.\ RAG},{RAPTOR},{Tree Structure},{\treekg{} (ours)}},
    yticklabel style={font=\scriptsize},
    bar width=9pt, bar shift=0pt,
    nodes near coords, nodes near coords style={font=\tiny},
    every node near coord/.append style={/pgf/number format/fixed,
        /pgf/number format/precision=2, /pgf/number format/print sign},
    axis x line*=bottom, axis y line*=left,
    enlarge y limits=0.22,
    xmajorgrids, grid style={inkG!18},
]
\addplot[draw=inkE!80, fill=inkE!30, bar shift=0pt] coordinates {(-18.57,image)};
\addplot[draw=inkE!80, fill=inkE!30, bar shift=0pt] coordinates {(-8.54,raptor)};
\addplot[draw=inkB!80, fill=inkB!30, bar shift=0pt] coordinates {(2.15,tree)};
\addplot[draw=inkA!85, fill=inkA!45, bar shift=0pt] coordinates {(6.43,treekg)};
\end{axis}
\end{tikzpicture}%
}
\caption{Top-1 accuracy relative to the text-only RAG baseline. Preserving the
native hierarchy is worth $+2.15$ points; adding per-guide knowledge-graph
entity matching on top is worth a further $+6.43$, for a total of $+8.58$.
RAPTOR's \emph{constructed} hierarchy and caption-and-inject multimodal
enrichment both fall below the flat baseline.}
\label{fig:delta}
\end{figure}
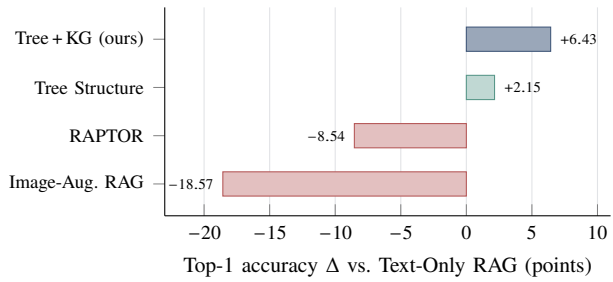

\treekg{} achieves the highest accuracy at every cutoff: \textbf{54.78\%} at
Top-1, \textbf{77.07\%} at Top-3, and \textbf{82.48\%} at Top-5. The Top-1
result is the most informative. It is an \textbf{8.58}-point improvement over
the text-only baseline and \textbf{6.43} points over tree-structured retrieval
without the knowledge graph, which isolates the contribution of entity matching
from the contribution of structure
(Figure~\ref{fig:delta}).

The decomposition in Figure~\ref{fig:delta} answers RQ1 and RQ2 separately, and
the two answers differ in strength. Preserving native structure with problem
abstractions is worth $+2.15$ Top-1 points over flat chunking---a real but
modest effect that, on a corpus of this size, corresponds to roughly seven
additional incidents and is within sampling variation
(Section~\ref{sec:limitations}). Adding per-guide entity matching on top is
worth a further $+6.43$ points, roughly three times as large and well outside
that range. Entity matching is therefore not a marginal refinement; it supplies
the decisive signal precisely when the embedding similarity margin between
competing guides is small, which is the regime that determines Top-1.

The gap between Top-1 and Top-5 for \treekg{} is 27.70 points, which carries a
direct operational implication: presenting a short ranked list of five
candidates rather than a single suggestion substantially increases the
likelihood of surfacing the correct guide, at negligible cost to an engineer who
is going to skim them in seconds.

\begin{keyfinding}[title={RQ1 + RQ2}]
Structure and entities contribute independently and unequally. Native hierarchy
with problem abstractions buys $+2.15$ Top-1 points over flat chunking; adding
per-guide knowledge-graph entity matching buys a further $+6.43$---about three
times as much, and the only one of the two large enough to be resolved on a
corpus of this size. The discrete signal matters most exactly where the
continuous one is ambiguous.
\end{keyfinding}

\subsection{Native Structure Beats Constructed Structure}
\label{sec:res_structure}

The comparison between the two hierarchical methods is the cleanest
architectural result in the study. RAPTOR, which \emph{constructs} a hierarchy
by recursively clustering and summarizing chunks, reaches only 37.66\% Top-1 and
72.08\% Top-5---below the flat text-only baseline at every cutoff. Our
tree-structured method, which \emph{reads} the hierarchy the author already
wrote, reaches 48.35\% Top-1, above the flat baseline.

We attribute the difference to what each abstraction step preserves. RAPTOR's
recursive summarization is lossy in a way that is harmful here: it merges
sibling content into a higher-level summary, and the discriminative detail that
separates two similar guides---a specific error code, a specific scale-unit
behavior---is exactly the kind of detail summarization discards. Native-%
hierarchy nodes, by contrast, are abstracted \emph{into problem statements}
rather than \emph{into summaries}, which changes the register without averaging
away the specifics.

An honest caveat belongs here. Tree Structure trails the text-only baseline at
Top-5 (75.68\% vs.\ 81.20\%), even though it leads at Top-1 ($+2.15$) and
essentially ties at Top-3 ($+0.17$). Structure sharpens the head of the ranking
and slightly narrows the tail: focusing each embedding on a specific sub-problem
improves precise discrimination but makes the method less able to surface a
loosely related guide as a fifth-place fallback. Adding the knowledge graph
resolves this---\treekg{} leads at Top-5 as well (82.48\%)---because entity
overlap recovers guides that share concrete entities with the incident even when
no single node abstraction is a close embedding match.

\subsection{Visual Understanding Does Not Help---It Hurts (RQ3)}
\label{sec:res_visual}

We expected image-augmented retrieval to help. It does the opposite, and by a
wide margin: 27.63\% Top-1 and 58.56\% Top-5, which is \textbf{22.64 points
below} the text-only baseline at Top-5 and 18.57 points below at Top-1. This is
the largest single effect in the study and the only one with a negative sign.

Our error analysis identifies three compounding causes:

\begin{enumerate}
  \item \textbf{Caption genericity.} VLM captions were frequently too generic to
        discriminate---a description amounting to ``a screenshot of a monitoring
        dashboard'' is true of a large fraction of the corpus and therefore
        carries no retrieval signal at all, while still consuming embedding
        capacity.
  \item \textbf{Amplified dilution.} Injecting caption text into guides that
        were already long worsens exactly the embedding-dilution problem that
        structure-aware methods are designed to fix. The intervention makes the
        core failure mode worse.
  \item \textbf{Redundancy.} Much image content restates the surrounding prose.
        The author pasted a screenshot of the query output \emph{and} described
        what it shows, so the caption adds tokens without adding information.
\end{enumerate}

\begin{keyfinding}[title={RQ3 --- negative result}]
For retrieval, caption-and-inject multimodal enrichment is not neutral, it is
harmful: $-22.64$ Top-5 points. Generic captions dilute embeddings while adding
no discriminative signal. This is a caution about the caption-and-inject
pattern, not about visual understanding in general---native visual retrieval
that never reduces an image to a caption~\cite{faysse2024colpali,cho2024m3docrag}
is a different design point we did not evaluate.
\end{keyfinding}

Taken together with Section~\ref{sec:res_structure}, the picture is consistent:
retrieval accuracy in this domain is far more sensitive to \emph{how content is
structured} than to \emph{how much content is added}. Both of our
negative results---RAPTOR and image augmentation---are cases where a method
added abstraction or content at the cost of discriminative precision.

\subsection{Near-Duplicate Guide Detection}
\label{sec:res_similarity}

An ancillary capability emerged from the same representation. Because \treekg{}
represents every guide as a set of problem abstractions plus an entity graph, it
can identify guides that are extremely similar to one another---cases where
multiple guides address overlapping or nearly identical problems. These are
precisely the cases that produce retrieval confusion, and they are also a
maintenance liability: an engineer who finds one of a pair of near-duplicate
guides has no way to know the other exists and may be more current. The system
therefore supports repository hygiene by flagging consolidation and rewrite
candidates, converting a retrieval difficulty into an actionable
documentation-quality signal.

\section{Analysis}
\label{sec:analysis}

\begin{figure}[t]
\centering
\resizebox{0.92\columnwidth}{!}{%
\begin{tikzpicture}[
    >={Stealth[length=2.5mm]},
    node distance=0.35cm,
    mbox/.style={rectangle, rounded corners=2pt, draw=#1, fill=#1!12,
                 minimum width=3.0cm, minimum height=0.68cm,
                 align=center, font=\small\sffamily},
    lbl/.style={font=\small\sffamily\bfseries, color=#1},
]
\node[lbl=inkC] (l1) {Text-Only RAG};
\node[mbox=inkC, below=0.22cm of l1] (flat) {Fixed-size flat chunks};
\node[mbox=inkC, below=of flat] (avg) {Averaged embedding};
\node[mbox=inkE, below=of avg] (dilute) {Diluted signal};
\draw[->, semithick, inkC!75] (flat) -- (avg);
\draw[->, semithick, inkE!70] (avg) -- (dilute);

\node[lbl=inkA, right=2.6cm of l1] (l2) {\treekg{} (ours)};
\node[mbox=inkA, below=0.22cm of l2] (tree) {Native hierarchy nodes};
\node[mbox=inkB, below=of tree] (prob) {Problem abstractions};
\node[mbox=inkD, below=of prob] (kg) {KG entity matching};
\draw[->, semithick, inkA!75] (tree) -- (prob);
\draw[->, semithick, inkD!75] (prob) -- (kg);

\node[font=\footnotesize\sffamily, color=inkG] at ($(avg.east)!0.5!(prob.west)$)
  {vs.};
\end{tikzpicture}%
}
\caption{Failure modes compared. Flat retrieval averages a long, multi-topic
guide into one diluted vector. \treekg{} decomposes the guide along its own
section boundaries, rewrites each subtree into problem-oriented language, and
adds a discrete entity signal that cosine similarity cannot express.}
\label{fig:comparison}
\end{figure}
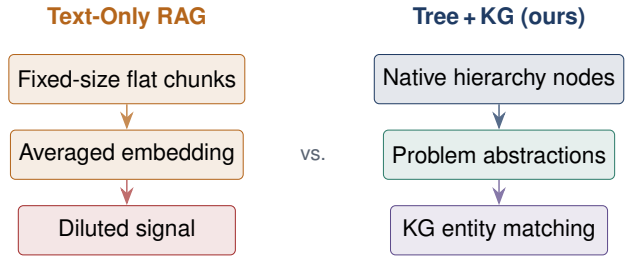

\subsection{Why \treekg{} Outperforms Flat Retrieval}
\label{sec:analysis_why}

The primary failure mode of text-only RAG is \emph{embedding dilution}: long,
multi-section guides produce averaged representations that lose discriminative
detail (Figure~\ref{fig:comparison}). \treekg{} attacks this at three levels.

\textbf{Structural decomposition} splits each guide along its own authored
boundaries into nodes that each concern one sub-problem, and the $\max$ over
nodes in the scoring function means a guide is judged by its best-matching
sub-problem rather than by an average over everything it covers.

\textbf{Problem abstraction} closes the vocabulary gap. This is the component
that makes the structure useful: decomposing a guide into sections that are
still written in solution language would produce many small embeddings that are
all equally far from the incident. Rewriting each subtree into a problem
statement moves the embeddings into the query's region of the space.

\textbf{Entity-level matching} supplies signal of a different kind. Service
names, error codes, and diagnostic tool names are discrete, high-precision
tokens; two guides can be nearly indistinguishable under cosine similarity while
differing sharply in which error codes they mention. Continuous similarity
cannot express ``this exact code appears here and nowhere else,'' and that is
exactly what the graph contributes.

\subsection{Why Visual Enrichment Underperformed}
\label{sec:analysis_visual}

The failure of image-augmented retrieval is instructive precisely because the
motivating hypothesis was reasonable. Guides really do contain diagnostically
important screenshots and diagrams; a human reading the guide really does use
them. The mistake was assuming that information useful \emph{to a reader who has
already found the guide} is also useful \emph{for finding it}.

Retrieval rewards discriminative tokens, and generic captions are the opposite of
discriminative. Worse, captioning has an asymmetric cost profile: a caption that
adds no signal still adds length, and length is the mechanism behind the
dilution failure mode that dominates this task. The result is a method that pays
the full cost of the intervention and collects none of the benefit.

We emphasize the scope of this finding. Visual understanding may well be
valuable for \emph{downstream} tasks---postmortem generation, guided execution,
human-assisted diagnosis---where an image is consumed rather than summarized. It
is specifically the retrieval stage, under the caption-and-inject pattern, where
structural approaches dominate. Native visual retrieval that embeds page images
directly~\cite{faysse2024colpali,cho2024m3docrag} avoids the captioning
bottleneck entirely and is the more promising multimodal direction for future
work.

\section{Discussion}
\label{sec:discussion}

\subsection{Deployment Recommendations}

Three recommendations follow directly from the results.

\textbf{Return five candidates, not one.} The gain from Top-1 to Top-5 is 27.70
points. Presenting a single suggestion discards more than a quarter of the
system's achievable value, and an OCE can dismiss four wrong candidates in
seconds.

\textbf{Make \treekg{} the retrieval backbone.} Knowledge-graph construction is a
one-time offline cost per guide, incurred when a guide is authored or edited and
never on the incident critical path, while the Top-1 gain of 8.58 points over
the flat baseline is realized on every incident. The economics are decisively
favorable, and the per-guide (rather than corpus-wide) graph design keeps
re-indexing an edited guide cheap.

\textbf{Use similarity detection proactively.} Near-duplicate guides degrade both
human and machine retrieval. The same representation that powers recommendation
identifies them, and acting on that signal improves the corpus rather than only
navigating around its defects.

\subsection{Limitations}
\label{sec:limitations}

Our evaluation covers 314 incidents, which bounds the resolution of the
comparisons we can make. Treating each Top-$k$ accuracy as a binomial
proportion, a single method's accuracy near 50\% carries a 95\% interval of
roughly $\pm 5.5$ points at this sample size. The large effects we report are
comfortably outside that range---\treekg{}'s $+8.58$ Top-1 gain over the flat
baseline, and the $-22.64$ Top-5 penalty from image augmentation---and the
method ordering at Top-1 is stable. The $+2.15$ margin by which tree-structured
retrieval alone exceeds flat chunking is not: it amounts to about seven
incidents and should be read as suggestive rather than established. We note
that all five methods were run over the identical incident set, so the
comparisons are paired and a paired test would have more power than the
unpaired bound above; we report the conservative figure because we did not
retain per-incident outcomes needed for a paired analysis.

Our ground truth is the guide label recorded in the incident-management system,
which reflects the guide an engineer \emph{used}, not necessarily the best guide
available; a system that recommends a better guide than the one used is scored
as wrong. Knowledge-graph construction depends on LLM entity extraction and can
introduce spurious entities, which we did not separately quantify. The current
matching function uses simple normalized entity overlap; graph-aware similarity
that exploits relation types and paths is unexplored and is a natural source of
further gains. The corpus comes from a single organization's guide repository
and incident-management toolchain, and both guide-authoring conventions and the
severity of the vocabulary gap will vary elsewhere. Finally, consistent with
Section~\ref{sec:privacy}, all results are corpus-level aggregates; we do not
report per-team or per-guide breakdowns, so we cannot characterize variance
across the 18 teams.

\subsection{Future Work}
\label{sec:future}

Four directions follow. \emph{Merging per-guide graphs} into a unified graph
would enable cross-guide reasoning---identifying that two guides address
different stages of the same failure---at the cost of the boundary sharpness that
currently helps us; the graph-RAG literature offers well-developed machinery for
this~\cite{edge2024graphrag,guo2024lightrag,gutierrez2024hipporag,peng2024graphragsurvey}.
\emph{Graph-aware similarity} that scores relation paths rather than entity
overlap is the most direct extension of our scoring function.
\emph{Judge-based evaluation} would relax the assumption that the recorded guide
is the correct one, and would let us credit a recommendation that is better than
the label~\cite{zheng2023judging,es2024ragas,saadfalcon2024ares}.
\emph{Native visual retrieval} in the ColPali style~\cite{faysse2024colpali}
would test whether guide images carry retrievable signal when they are not first
collapsed into captions---the one version of RQ3 our negative result does not
answer. Finally, integration with guide \emph{execution}
agents~\cite{an2024nissist} would close the loop from retrieval to mitigation.

\section{Conclusion}
\label{sec:conclusion}

We presented \system{}, a retrieval system that recommends troubleshooting
guides during cloud incident response. Through a systematic evaluation of five
retrieval strategies on 314 real-world incidents spanning 112 guides and 18
service teams, we showed that our \treekg{} method achieves the highest accuracy
at every cutoff (54.78\% Top-1, 77.07\% Top-3, 82.48\% Top-5), outperforming
flat, recursive-abstractive, and image-augmented baselines.

The central lesson is that \emph{structural alignment} beats semantic
enrichment. Preserving the hierarchy the guide's author already wrote, rewriting
each subtree into the problem-oriented register that incidents are phrased in,
and adding a discrete per-guide entity signal together contribute 8.58 Top-1
points over a flat baseline. Two interventions that add abstraction or content
instead of structure---RAPTOR's constructed hierarchy and caption-and-inject
multimodal enrichment---both fall below that baseline, the latter by a wide
margin. That concurrent work on hierarchically structured
documents~\cite{wang2025bookrag} independently arrives at the same
tree-plus-entity-graph recipe from a different starting point strengthens our
confidence that this is the right general architecture for retrieval over
authored technical documentation, and not an artifact of our corpus.

\balance

\end{document}